\documentclass{article}
\usepackage{arxiv}
\usepackage[utf8]{inputenc} % allow utf-8 input
\usepackage[T1]{fontenc}    % use 8-bit T1 fonts
\usepackage{hyperref}       % hyperlinks
\usepackage{url}            % simple URL typesetting
\usepackage{booktabs}       % professional-quality tables
\usepackage{amsfonts}       % blackboard math symbols
\usepackage{nicefrac}       % compact symbols for 1/2, etc.
\usepackage{microtype}      % microtypography
\usepackage{lipsum}
\usepackage{graphicx}
\usepackage{amsmath}
\usepackage{nccmath}
\usepackage[super]{nth}
\graphicspath{ {./images/} }

\title{Early Indicators of COVID-19 Spread Risk Using Digital Trace Data of Population Activities with Different Purpose}

\author{
 Xinyu Gao\thanks{Corresponding author $^+$ Xinyu Gao and Chao Fan have equal contribution to this paper.}  \space $^+$ \\
  Zachry Department of Civil and \\Environmental Engineering\\
  Texas A\&M University\\
  College Station, Texas, 77843 \\
  \texttt{xy.gao@tamu.edu} \\
  \And
 Chao Fan$^+$\\
  Zachry Department of Civil and \\Environmental Engineering\\
  Texas A\&M University\\
  College Station, Texas, 77843 \\
  \texttt{chfan@tamu.edu} \\
  \And
 Yang Yang \\
  Department of Computer Science\\ \& Engineering\\
  Texas A\&M University\\
  College Station, Texas, 77843 \\
  \texttt{yangyangsandy@tamu.edu} \\
  \And
 Sanghyeon Lee \\
  Department of Electrical and \\Computer Engineering\\
  Texas A\&M University\\
  College Station, Texas, 77843 \\
  \texttt{sanghyeonlee@tamu.edu} \\
  \And
 Qingchun Li \\
  Zachry Department of Civil and \\Environmental Engineering\\
  Texas A\&M University\\
  College Station, Texas, 77843 \\
  \texttt{qingchunlea@tamu.edu} \\
  \And
  Mikel Maron \\
  Community Team\\
  Mapbox\\
  Washington, DC, 20005 \\
  \texttt{mikel@mapbox.com} \\
  \And
 Ali Mostafavi* \\
  Zachry Department of Civil and \\Environmental Engineering\\
  Texas A\&M University\\
  College Station, Texas, 77843 \\
  \texttt{mostafavi@tamu.edu} \\
}

\begin{document}
\maketitle 
\begin{abstract}

The spread of pandemics such as COVID-19 is strongly linked to human activities. The objective of this paper is to specify and examine early indicators of disease spread risk in cities during the initial stages of outbreak based on patterns of human activities obtained from digital trace data. In this study, the \textit{Venables distance} ($D_v$), and the \textit{activity density} ($D_a$) are used to quantify and evaluate human activities for 193 US counties, whose cumulative number of confirmed cases was greater than 100 as of March 31, 2020. Venables distance provides a measure of the agglomeration of the level of human activities based on the average distance of human activities across a city or a county (less distance could lead to a greater contact risk). Activity density provides a measure of level of overall activity level in a county or a city (more activity could lead to a greater risk). Accordingly, Pearson correlation analysis is used to examine the relationship between the two human activity indicators and the basic reproduction number in the following weeks. The results show statistically significant correlations between the indicators of human activities and the basic reproduction number in all counties, as well as a significant leader-follower relationship (time lag) between them. The results also show one to two weeks’ lag between the change in activity indicators and the decrease in the basic reproduction number. This result implies that the human activity indicators provide effective early indicators for the spread risk of the pandemic during the early stages of the outbreak. Hence, the results could be used by the authorities to proactively assess the risk of disease spread by monitoring the daily Venables distance and activity density in a proactive manner.

\end{abstract}

\section{Introduction}
The objective of this study is to reveal and evaluate early indicators for COVID-19 spread risk in cities during the initial stages of the outbreak using measures of human activities derived from digital trace data. An arguably unprecedented global pandemic, the coronavirus disease 2019 (COVID-19) has infected millions of people worldwide with a mortality rate of 6.6\% and a high infection rate \cite{WHO}\cite{keni2020covid}. Since the spread of COVID-19 is highly dependent on human activities, incidence of infection could be contained by restricting human activities and mobility \cite{gollwitzer2020connecting}. Many countries and authorities have implemented various non-pharmaceutical interventions (e.g., shelter-in-place orders, regional lockdowns, and travel restrictions), which were undertaken to slow the spread of disease by disrupting transmission chains by restricting human mobility and activities. Such social distancing and activity reduction interventions have proven to be critical in slowing down the spread of pandemics both in previous epidemics \cite{caley2008quantifying} and during COVID-19 \cite{anderson2020will}\cite{tian2020investigation}\cite{ramchandani2020deepcovidnet}.

While reduction in human activities is considered an effective measure for containing epidemics and pandemics, there are limited reliable, proven, real-time leading indicators related to human activities that could provide early insights about the risk of disease spread in a region to inform proactive policy making. One reason for this limitation has been the absence of quantitative measures and data that could be examined to proactively evaluate human activities. With advancements in location intelligence data technologies, however, information derived from cellular devices offers a large depository of digital trace data related to human activities have been have increasingly been adapted and analyzed to promote understanding of and to quantify human activity and mobility in pandemic analysis, as well as in other applications \cite{asgari2013survey}\cite{balcan2009multiscale}\cite{barbosa2018human}. For example, in the context of COVID-19, the radius of gyration, which captures the mobility of individuals using human movement trajectories, was adopted to analyze the COVID-19 spread in Japan \cite{yabe2020non}. Daily step-counts (gathered from smartphones) were used to estimate and predict decreased movement of individuals within the United States during COVID-19 \cite{gollwitzer2020connecting}. Two of the most important aspects of human activities during an epidemic are agglomeration of activities and intensity of activities.

Although previous research reveals insights regarding human activities in the context of COVID-19, the relationship between human activities and disease-spreading risk has not been fully explored, and leading indicators of human activities to proactively assess the risk of disease spread during the early stages of pandemics are lacking. The majority of research studies \cite{chang2020mobility}\cite{cintia2020relationship}\cite{gao2020mapping}\cite{li2020disparate} focus on quantifying and analyzing the changes in human activities as a consequence of the outbreak of the virus and in response to protective policies (such as shelter-in-place policies). The time-lag relationship between these human activity metrics and the spread of virus, which can be generally described by the basic reproduction number ($R_0$), has not yet been fully examined. The basic reproduction number, $R_0$, is defined as the number of secondary cases produced by one previous case in a completely susceptible population \cite{dietz1993estimation}. Although research studies \cite{lampos2020tracking}\cite{lu2020internet} have focused on leading indicators obtained from users’ online search behavior, the decrease of online search frequencies may not have direct impact on the spread of virus. Hence, the previous indicators cannot be utilized for proactive assessment of disease spread risk in a proactive manner. 

In addition, the nature of human activities—such as activities in public indoor venues versus in residences—may have differing effects on the spread risk of a disease. The measures of human activities should distinguish between the nature of activities to provide useful leading insights for decision making and policy formulation.

In this study, we adopted the Venables distance ($D_v$) index \cite{louail2014mobile} and also created the activity density ($D_a$) index to serve as two real-time indicators to examine the spatial and temporal patterns of human activities across 193 counties in the United States using Mapbox high-resolution temporal-spatial activity index data from January 1 to March 31, 2020. The Venables distance captures the average distance (i.e., concentration) of human activities across a city or county (less distance between persons might indicate a greater contact risk). The activity density captures the intensity level of overall activities in a county or city (higher activity levels might indicate a greater spread risk). Human activities were examined in four categories—social, traffic, work, and other—based on the location and time of activities. Accordingly, we analyzed the correlation between the two metrics ($D_v$ and $D_a$) and the basic reproduction number for 193 counties with the highest number of confirmed COVID-19 cases. The rest of this paper is organized into three sections. The first section discusses the description of the two datasets (Mapbox data and total confirmed cases number data), as well as the analysis methods. The second section describes the results of time-lag correlation analysis between the two metrics and the basic reproduction number. The last section presents the results and the implications of the findings for future work.

\section{Methods}
In this section, we describe the two datasets—Mapbox data and total confirmed cases number data—and the procedures for human activity categorization. Also covered in this section are definitions and equations related to the Venables distance ($D_v$), the activity density ($D_a$), and the basic reproduction number ($R_0$). The time-lag cross-correlation analysis method is presented at the end of this section. 

\subsection{Data Source and Preprocessing}
\begin{figure}[!ht]
\centering
\includegraphics[width=\linewidth]{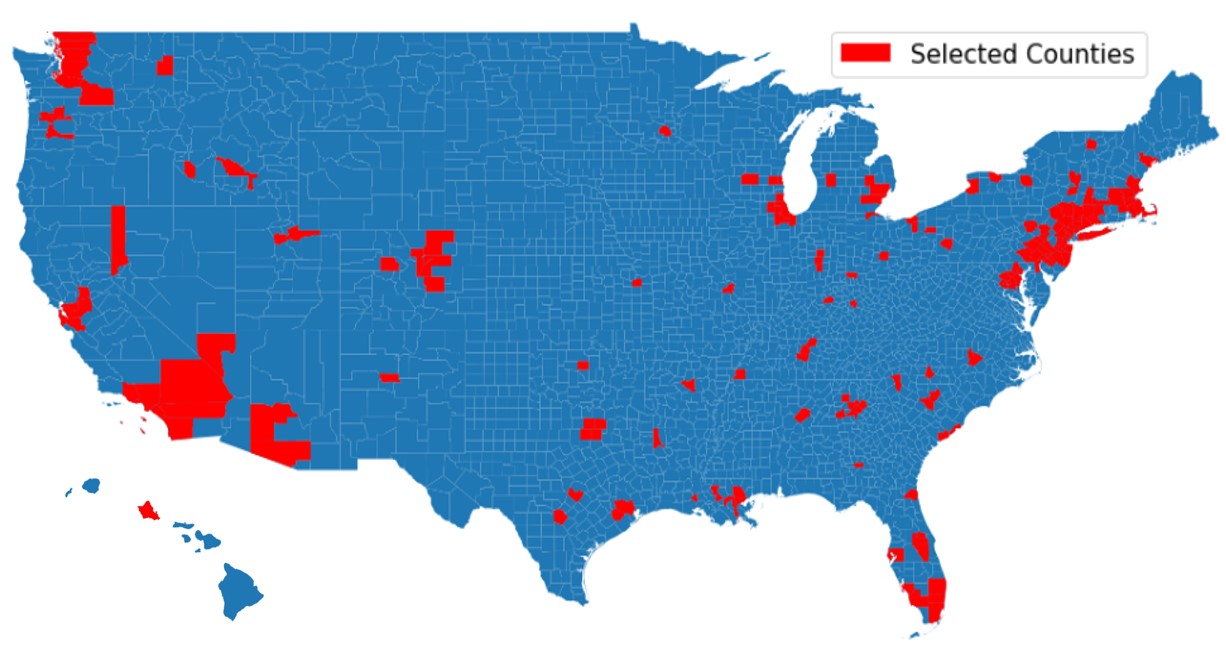}
\caption{193 selected counties whose cumulative confirmed cases were greater than 100 as of March 31, 2020.}
\label{fig:1}
\end{figure}

We utilized digital trace telemetry data obtained from Mapbox from January 1 to March 30, 2020. The dataset contains a metric of telemetry-based human activity, $a_{tile,t}$, which varies across spatial tiles and time t. The partition of tiles is based on Mercantile, a Python library, which is capable of creating spatial-resolution grids all worldwide. The $a_{tile,t}$ is collected, aggregated, and normalized by Mapbox from geography information updates from users’ cell phone locations by time flows. The more users located in a tile at time t, the higher the human activity (i.e., $a_{tile,t}$). The dataset comprises the United States and the District of Columbia; however, in this study, we examined only 193 counties whose cumulative confirmed cases were greater than 100 as of March 31, 2020 (shown in \ref{fig:1}). In the raw data, the temporal resolution is 4 hours. Each tile represents about 100 by 100 square meters for spatial resolution. Since the data is derived from cell phone activity, data may not exist for all cells at all times. For example, a park opens during the day but is closed at night would not generate any data at midnight. Also, for protecting users’ privacy and the data aggregation process, tiles with a small number of users are reported without any activity data. It is also noteworthy that the data is aggregated and normalized for each month, so the absolute values of activity indices for different months cannot be directly compared.
To reveal the time-lag relationship between metrics and spread of the virus, the total number of confirmed cases was used. We obtained the data from the COVID-19 Data Repository by the Center for Systems Science and Engineering (CSSE) at Johns Hopkins University \cite{Hopkins}. The data in this repository were gathered and aggregated from various sources, such as the World Health Organization (WHO) and the Centers for Disease Control and Prevention (CDC).  We extracted the total number of confirmed cases $c_{i,j}$ from the CSSE repository, where i represents each date and j represents each county.

\subsection{Tile Categorization}
\begin{figure}[!ht]
\centering
\includegraphics[width=0.7\linewidth]{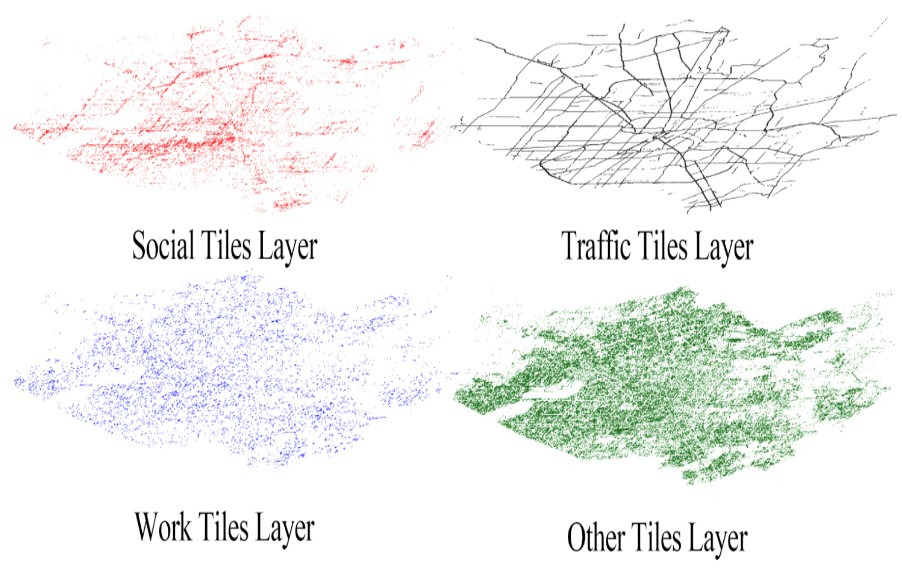}
\caption{Maps of four different tile categories (Social, Traffic, Work, and Other) in Harris County, Texas.}
\label{fig:2}
\end{figure}

The nature of an activity might put its participants at a higher risk level for contracting the virus. For example, activities in public common areas, such as grocery stores or gyms, would lead to greater risk of disease spread compared to the activities in residential areas, such as working from home or walking a dog in the community. The fine granularity of the spatial resolution enables classification of each tile into one of the four categories: (1) social tiles, (2) traffic tiles, (3) work tiles, and (4) other tiles. Categorization is based on the following characteristics: (1) social tiles are the location of at least one point of interest; (2) traffic tiles include a road; (3) work tiles show no activity in the evening; and (4) other tiles are located in residential areas. The analysis in this paper examines human activities for social, work, and traffic tiles separately. Example tile maps related to each category are shown in \ref{fig:2} for Harris County, Texas.

\subsection{Venables Distance}

To quantify the agglomeration of human activities, we used the Venables distance ($D_v$) as a weighted average distance of human activities. The Venables Distance aggregates the spatial distribution of $a_{tile,t}$ in a county and captures the urban spatial structure of human activities (Louail et al. 2014). The $D_v$ is calculated using Equation \ref{eq:1}:

\begin{equation}
\label{eq:1}
D_v(t) = \frac{\sum_{tile_1 \neq tile_2}{a_{tile_1,t}\times a_{tile_2,t}\times d_{tile_1,tile_2}}}{\sum_{tile_1 \neq tile_2}{a_{tile_1,t}\times a_{tile_2,t}}}
\end{equation}

where, $a_{tile_1,t}$ and $a_{tile_2,t}$ are the metrics of human activities in $tile_1$ and $tile_2$ at time t, respectively, and $d_{tile_1,tile_2}$ ) is the distance from the centroids between these two tiles. In Harris County, Texas, there are more than 70K unique tiles, which makes it computational expensive to analyze all pairs of existing tiles. To reduce the computational burden, we aggregated the 100 by 100-square-meter tiles, $a_{tile,t}$, to square cells 2 kilometers in length using Equation \ref{eq:2}:

\begin{equation}
\label{eq:2}
a_{k,t} = \frac{\sum_{\textrm{for all } tile \textrm{ in cell } k}{a_{tile,t}}}{A_k}
\end{equation}

where, $a_{k,t}$ is the intensity of human activity in cell $k$, at time $t$, $A_k$ is the area of the cell $k$. By aggregating human activity into a larger spatial cell, we reduced the computational efforts, maintaining a meaningful spatial resolution without losing important characteristics in the raw data. Accordingly, the modified Venables distance is derived as shown in Equation \ref{eq:3}:

\begin{equation}
\label{eq:3}
D_v(t) = \frac{\sum_{k_1 \neq k_2}{a_{k_1,t}\times a_{k_2,t}\times d_{k_1,k_2}}}{\sum_{k_1 \neq k_2}{a_{k_1,t}\times a_{k_2,t}}}
\end{equation}

where, $a_{k_1,t}$ and $a_{k_2,t}$ are the intensity of human activities in cells $k_1$ and $k_2$ at time $t$, respectively, and $d_{k_1,k_2}$  is the distance from the centroids between these two cells. In Equation \ref{eq:3}, the values of the activity intensity ($a_{k,t}$) are used as weights to calculate a human activity-weighted distance for the whole area. In other words, the relative values of $a_{k,t}$ were used to examine changes in agglomeration of activities. We calculated $D_v(t)$ for each county $j$, which is denoted as $D_v(j,t)$  for all cells in the county $j$.

\subsection{Activity Density}
Although $D_v$ captures the agglomeration of human activities, the density of activities is also critical for examining population contact. To make the raw data ($a_{tile,t}$) from Mapbox comparable among different months, we de-normalized the activity index to the contact activity metric $ca_{tile,t}$ for each tile and each month (where $t$ is time). In the de-normalization process, the assumption was that, in each month, the minimum human activity intensity among tiles is the same. First, the tile with minimum human activity index ($a_{tile,t}$) in each month is found. Then we set the value of contact activity $ca_{tile,t}$ in that low-activity tile as 5, and de-normalized the values for other tiles based on this value. For each county, the activity density at time $t$ ($D_a(t)$) was calculated using Equation \ref{eq:4}:

\begin{equation}
\label{eq:4}
D_a(t) = \sqrt{\frac{1}{N}\sum_{tile = 1}^{N}{ca_{tile,t}^{2}}}
\end{equation}

\subsection{Basic Reproduction Number Estimation}
The basic reproduction number ($R_0$) (the number of secondary cases arising from one previous case) is a critical parameter in epidemic modeling for understanding the speed of disease spread \cite{dietz1993estimation}\cite{gatto2020spread}, as well as the risk of virus spread \cite{aleta2020modeling}\cite{giordano2020modelling}\cite{liu2018measurability}\cite{newman2002spread}. $R_0$ is calculated using Equation \ref{eq:5}:

\begin{equation}
\label{eq:5}
c_{i+t} = c_i\cdot R_0^{t/\tau}
\end{equation}
where, $c_i$ and $c_{i+t}$ represent the total confirmed cases in day $i$ and day $i+t$, respectively; $\tau$ is a constant parameter. We estimated $R_0$ using CDC data ($c_{i,j}$) in Equation \ref{eq:6}:

\begin{equation}
\label{eq:6}
R_{0i,j} = e^{\tau\frac{\ln{c_{i,j}}-\ln{c_{i-t,j}}}{t}}
\end{equation}

where $R_{0i,j}$ is the basic reproduction number at date $i$ in county $j$. Because the CDC total confirmed cases data fluctuates within the course of a week (i.e., more reported cases in weekdays and less reported cases during weekends), the time interval $t$ was set to 7 days. Based on the existing literature and simulation models related to COVID-19 \cite{fan2020effects}\cite{zhang2020changes}\cite{zhang2020evolving}, the constant parameter $\tau$ was set to 5.1 days. Accordingly, the $R_0$ was calculated for all 193 counties for the analysis period.

\subsection{Time Lagged Cross-Correlation Analysis}
In the next step, we examined the correlation between the two human activity indicators and the basic reproduction number across all counties. Since these variables are a time series, we used time-lagged cross-correlation analysis to assess the synchrony of time series data sets. The cross-correlation coefficient was calculated using Equation \ref{eq:7}:

\begin{equation}
\label{eq:7}
\rho_{A_1,A_2}(\delta) = \frac{Cov(A_1(t),A_2(t+\delta))}{\sigma_{A_1}\sigma_{A_2}}
\end{equation}
where $\rho_{A_1,A_2}$ is the cross-correlation coefficient for two time series data $A_1$ and $A_2$; $\delta$ is the time offset of $A_2$; $Cov(X,Y)$ is the function calculating the covariance of two variables; $\sigma_{A_1}$ and $\sigma_{A_2}$ are the standard deviation of data $A_1$ and $A_2$, respectively. Based on the definition, $\rho_{A_1,A_2}$ represents the correlation between two variables and $|\rho_{A_1,A_2}|\leq1$ ( $|\rho_{A_1,A_2}|=1$ happens if and only if $A_1=mA_2+n$, where $m$ and $n$ are constants). Then, by iteratively calculating the $\rho_{A_1,A_2}(\delta)$ with different $\delta$, the correlation coefficient would reach its peak when $\delta=T$, and $T$ was determined as the time lag between two variables.

\section{Results}

This section presents the results related to the calculation of the two human activity metrics and their time-lagged correlation with the basic reproduction number across 193 counties during the initial stage of the COVID-19 outbreak in the United States.

\subsection{Evaluation of Human Activity for Each Category among Counties} 

\begin{figure}[!ht]
\centering
\includegraphics[width=\linewidth]{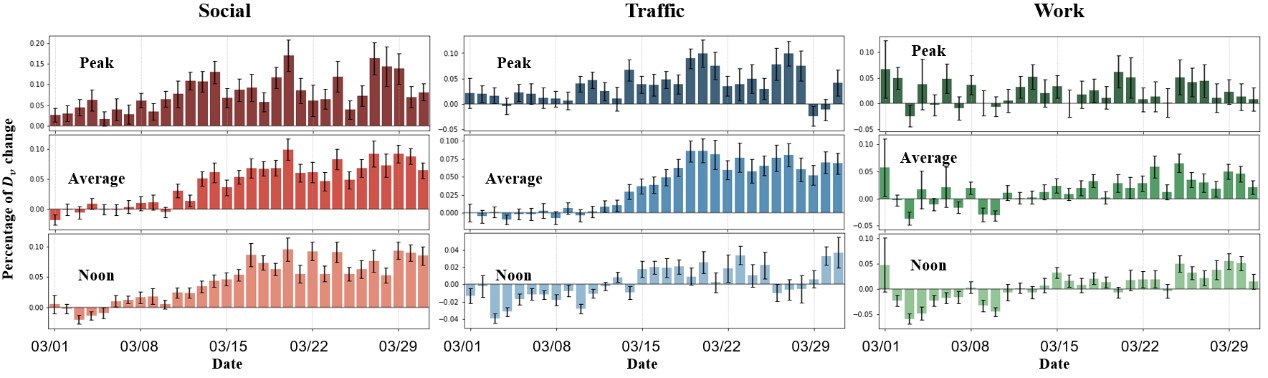}
\caption{The percentage change of $D_v$ for 193 counties in March 2020. The height of each bar is the average percentage change of all 193 counties. The error bar indicates the standard deviation among all counties. The three tile categories of social, traffic, and work are shown in each column, and three daily tile aggregations of peak, average, and noon are shown in each row.}
\label{fig:3}
\end{figure}

\begin{figure}[!ht]
\centering
\includegraphics[width=\linewidth]{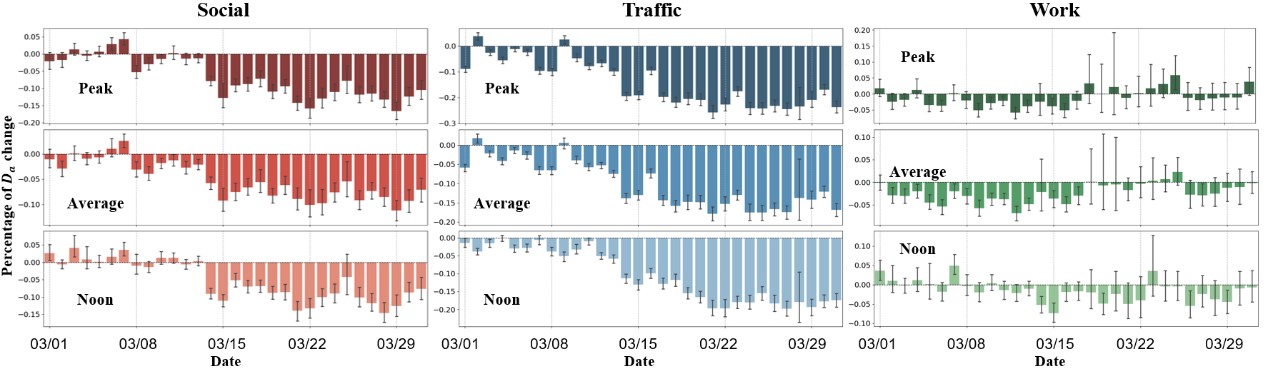}
\caption{The percentage change of $D_a$ for 193 counties in March 2020. The height of each bar is the average percentage change of all 193 counties, and the error bar indicates the standard deviation among all counties. The three tile categories of social, traffic, and work are shown in each column, and three daily tile aggregations of peak, average, and noon are shown in each row.}
\label{fig:4}
\end{figure}

In this study, the Venables distance ($D_v$), and the activity density ($D_a$) were calculated to assess the human activities at the county level using data from Mapbox. The very first four weeks (January 1 to 28) were considered as the baseline, and $D_v$ and $D_a$ values were compared with the average baseline in corresponding weekdays. For example, the $D_v$ values on March 1 (Sunday) were compared with the mean value of $D_v$ values on Sundays between January 1 and 28. Three different ways of daily activity aggregation were used: peak, average, and noon. The peak (largest), average, or the noon (11 a.m. to 3 p.m.) values of human activities $a_{tile,t}$ were selected and set as the representative value of each tile at each day. Figure \ref{fig:3} and Figure \ref{fig:4} show the percentage of $D_v$ and $D_a$ change for social, traffic, and work activity categories and for three types of daily tile aggregation. 

\begin{figure}[!ht]
\centering
\includegraphics[width=\linewidth]{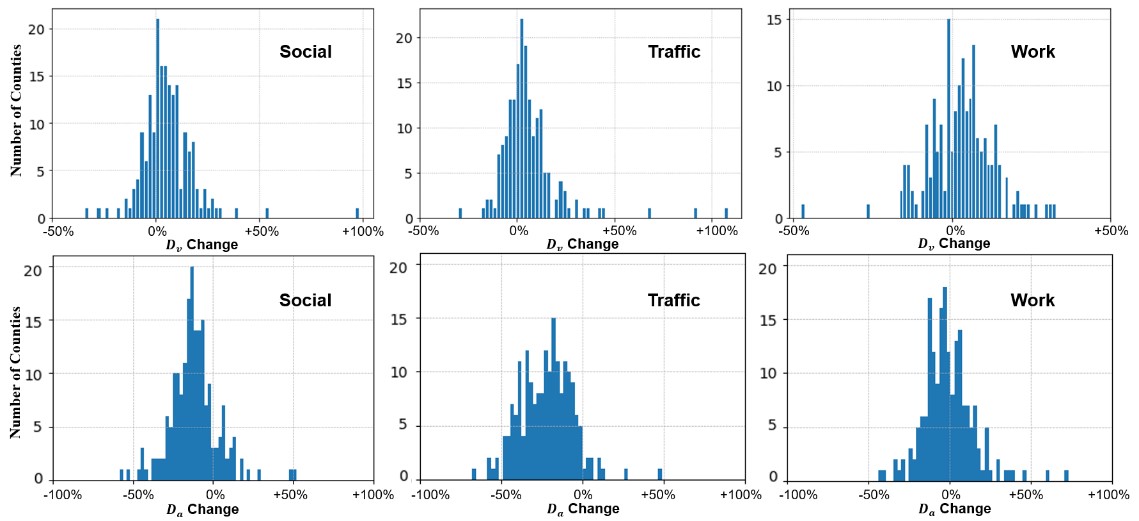}
\caption{Histogram plots of average percentage change of $D_v$ and $D_a$ (each row) for three different categories (each column).}
\label{fig:5}
\end{figure}

The increasing trend of Venables distance ($D_v$) implies declining concentration and rising distance among people, and the decreasing trend of activity density ($D_a$) indicates less human activity compared with the beginning of this year. Due to the shelter-in-place policies, residents changed their daily activity patterns. More and more people reduced the non-essential outdoor activities (e.g., shopping in supermarkets, exercising in gyms, and eating at restaurants). Such changes in daily human activity patterns led to the change of $D_v$ and $D_a$. For the three categories, significant change can be seen in both social and traffic tiles, while the change in work tiles is not obvious. This is because the activities in work tiles could be essential activities. The $D_v$ increased the most, around 15\%, in social tiles, while the $D_a$ fell the most in traffic tiles, which is around 25\%. Differences among the three types of daily tile aggregation were not significant. The percentage change of peak values is slightly greater than the other two values, indicating that peak values are influenced more by COVID-19, while average values are more stable. The following analysis uses peak values to calculate $D_v$ and $D_a$.
Histograms of average percentage change of $D_v$ and $D_a$ for each county are shown in Figure \ref{fig:5}. The average percentage change is calculated during the last week of March. The $D_v$ values in the majority of counties increased, and $D_a$ values decreased for social and traffic categories, while the work category shows more even distribution around 0\% for both $D_v$ and $D_a$. These histogram plots are consistent with the claim that human activities in work tiles are more essential than other two (social and traffic), which did not show significant change during the COVID-19 study period.

While $D_v$ and $D_a$ describe the different global characteristics of human activity—the $D_v$ captures spatial distribution of human activity, and the $D_a$ focuses on the intensity of human activity—they all reveal the insight of massive human activity patterns, which could have a quite significant influence to the spread of the virus. The correlation analysis between these two metrics and the basic reproduction number $R_0$ becomes critical.

\subsection{Time-Lagged Correlation Analysis}
\begin{figure}[!ht]
\centering
\includegraphics[width=\linewidth]{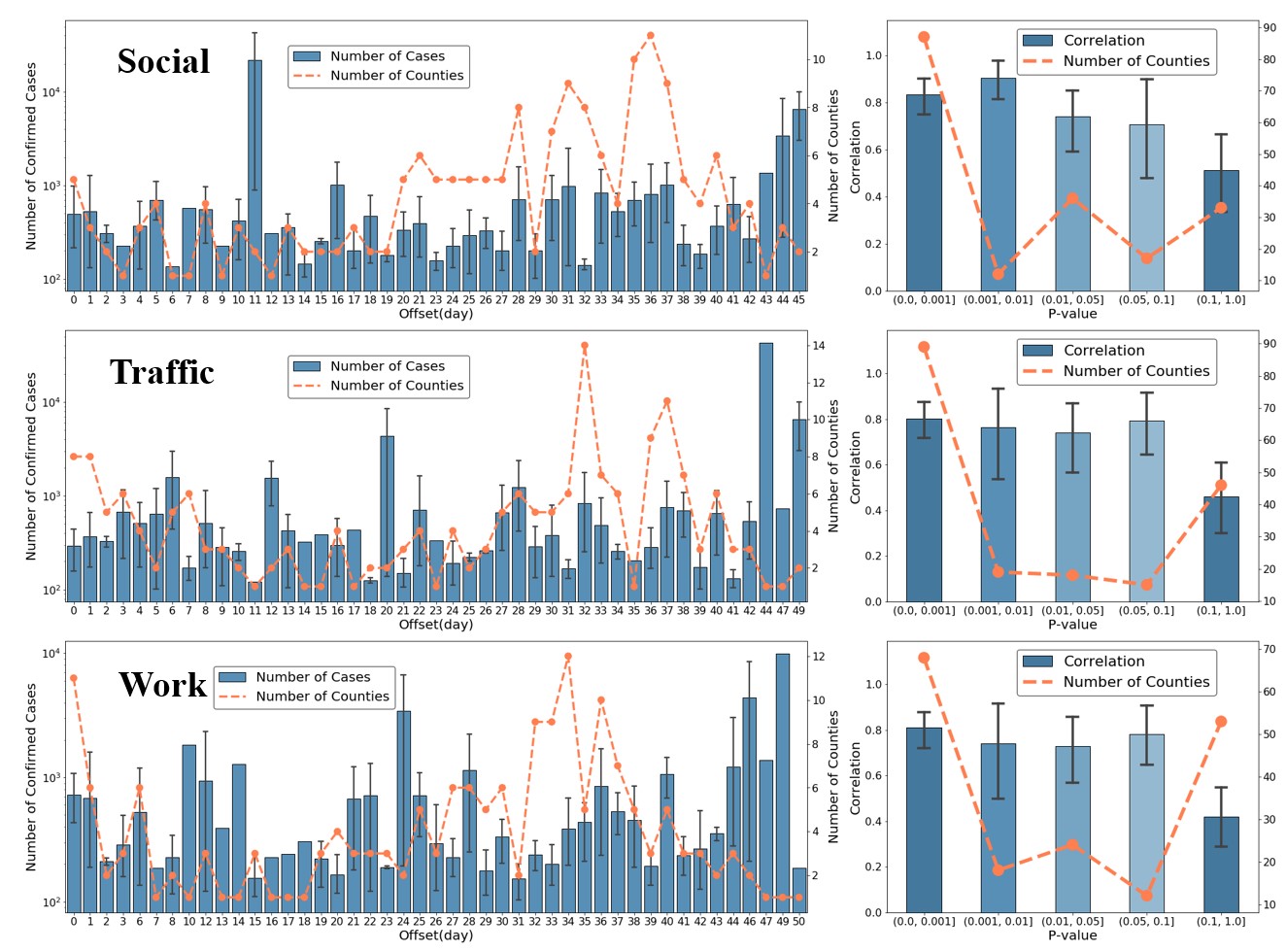}
\caption{The time-lagged correlation analysis between Venables distance ($D_v$) change and reproduction number ($R_0$). The left column shows the number of cases and the number of counties for different offset days, and the right column shows the correlation index and the number of counties with different P-value intervals. Each row presents one of the three tile activity categories: social, traffic and work.}
\label{fig:6}
\end{figure}

The spread of the coronavirus is closely related to the human activity patterns. In the previous section, we showed that the average distance between human activities ($D_v$) increased by 10\% to 15\%, and the average human activity intensity ($D_a$) decreased 5\% to 10\% for social tiles during March 2020 compared with the baseline period of January 2020. This result provides a good indication of the reduction in human activities in response to social distancing policies. In this section, we examine the extent to which the change in human activity metrics was related with the change in reproduction rate of corona virus in the 193 counties under study. Hence, we conducted the time-lagged correlation analysis for the two human activity metrics calculated based on social, traffic, and work tile activity categories. 
Figure \ref{fig:6} shows the time offset results between the Venables distance ($D_v$) change and the basic reproduction number ($R_0$) for social, traffic, and work activity categories. Since the number of confirmed cases follows a skewed distribution during March 2020, the log scale is used to illustrate the results. The results show that, in majority of typical counties, the decline in the basic reproduction number ($R_0$) happens 20 to 40 days after the increase in Venables distance. This result is consistent across all three activity categories. In the right column of Figure \ref{fig:6}, the bar charts show the correlation between the offset $D_v$ and $R_0$ within different P-value intervals. The average correlation coefficients (with P-values less than 0.05) are around 0.8 for each category, indicating a significant relationship between the increased distance among human activities and the decline in the virus spread speed. For the P-values greater than 0.05 (indicating no sufficient evidence to prove the correlation between two variables), the correlation indices are correspondingly smaller. The number of counties in each P-value interval show that about 50\% of the counties have P-values less than 0.001 for Venables Distance calculated based on social and traffic activity tiles. The results related to work tiles, however, show a significant correlation between the two variables in a smaller number of counties. 

\begin{figure}[!ht]
\centering
\includegraphics[width=\linewidth]{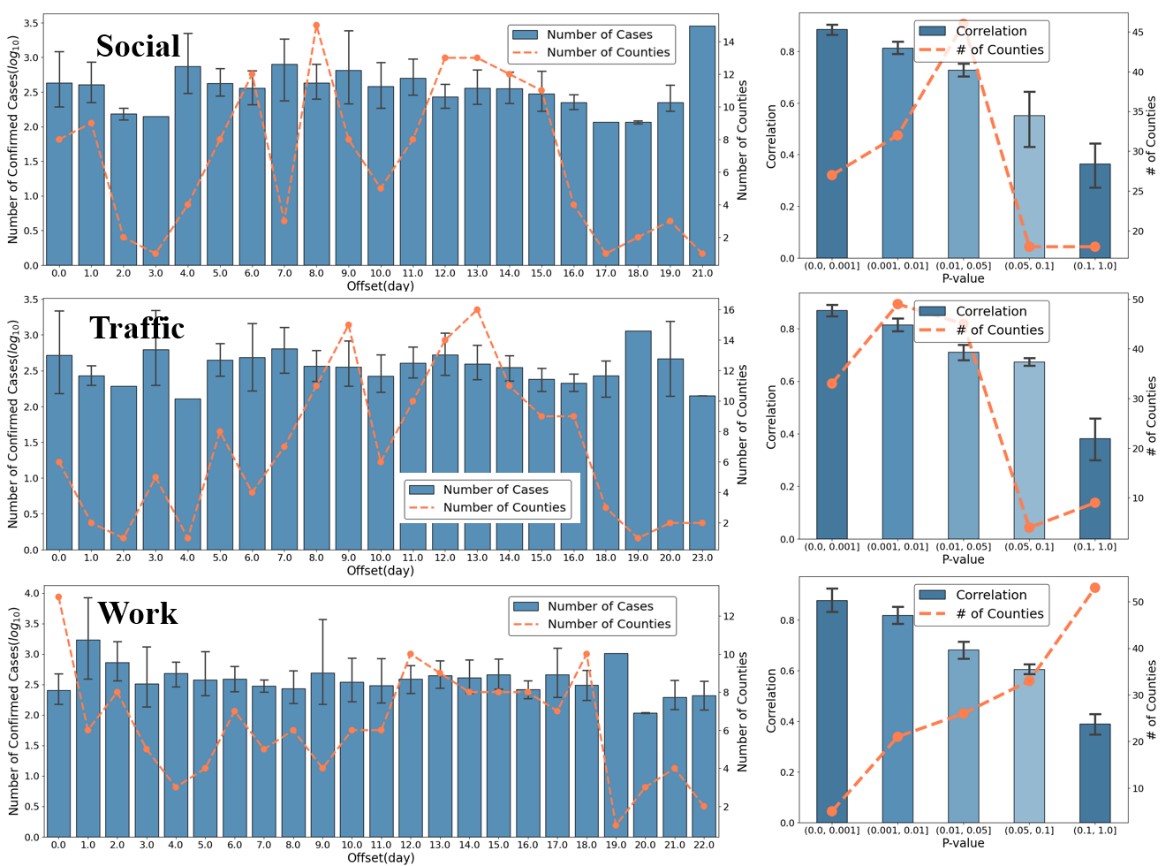}
\caption{The time-lagged correlation analysis result between activity density ($D_a$) change and reproduction number ($R_0$). The left column shows the number of cases and the number of counties for different offset days, and the right column shows the correlation index and the number of counties with different P-value intervals. Each row presents one of the three tile activity categories: social, traffic and work.}
\label{fig:7}
\end{figure}

Figure \ref{fig:7} shows the time offset result between activity density ($D_a$) change and reproduction number ($R_0$) for the three activity categories. The results show that the decline in the basic reproduction number happens 6 to 17 days after the reduction of the activity intensity ($D_a$); a similar result exists in all activity categories. The time lag is less than the one obtained for the Venables distance ($D_v$), which means that the spread of virus responds to human activity intensity reduction more quickly than to human agglomeration reduction. In the right column of Figure \ref{fig:7}, the bar charts show the correlation of offset $D_a$ and $R_0$ in different P-value intervals. The average correlation indices for P-values less than 0.05 are around 0.9 for tile activity categories. This result indicates a significant relationship between the human activity intensity reduction and the decline in the virus spread speed. For P-values greater than 0.05, the correlation indices are smaller as well. The number of counties in each P-value interval show about 50\% of counties have P-values less than 0.01 for social and traffic tiles, while the work tiles result shows P-values between 0.1 and 1.0 (indicating a non-significant correlation). 

\subsection{Heterogeneity for Different Features}

\begin{figure}[!ht]
\centering
\includegraphics[width=\linewidth]{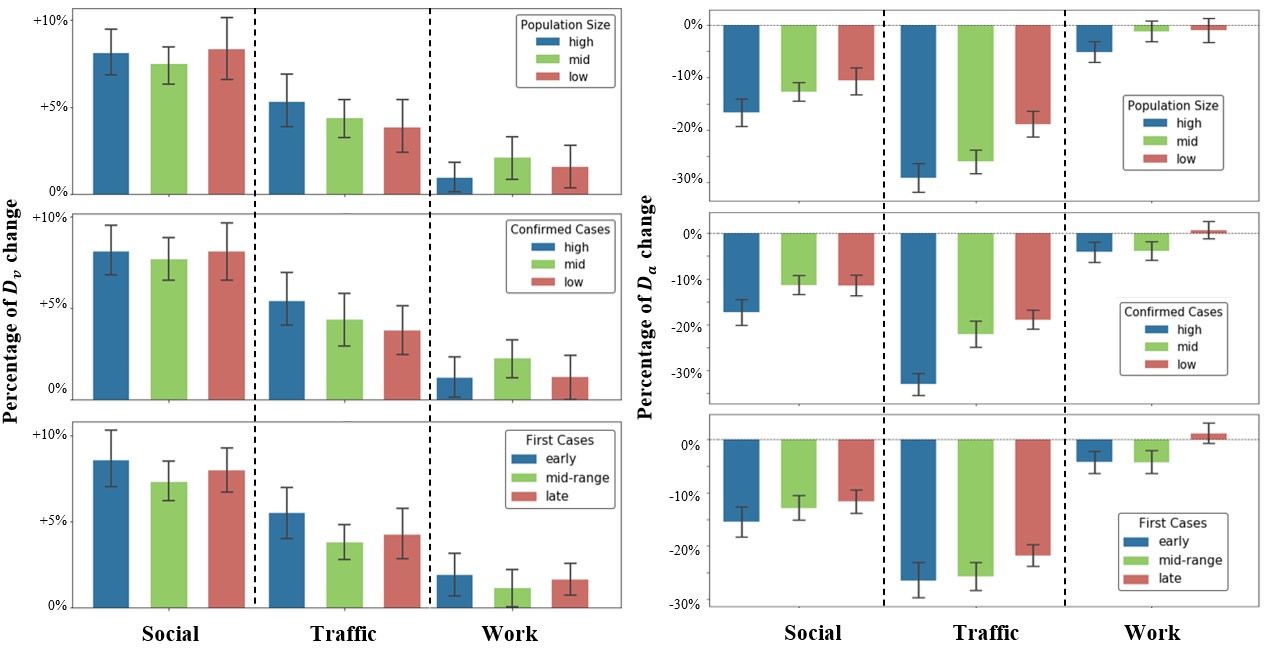}
\caption{. Change in Venables distances ($D_v$) and activity density ($D_a$) across counties with different population size, confirmed cases number, and the date of first confirmed case in the three tile activity categories (social, traffic, and work).}
\label{fig:8}
\end{figure}

In the next step, we examined the variation of findings across counties with different population sizes, number of confirmed cases, and date of first confirmed cases. The goal is to examine the extent to which the correlation between the two metrics of human activities and the reproduction number is sensitive to these county features. The 193 counties were divided into three uniform categories according to population size and confirmed cases (on March 18, 2020) labeled high, medium, and low. Similarly, the first case dates were labeled as early, mid-range, and late for each one third of counties. Then, the changes in $D_v$ and $D_a$ (on March 31, 2020) were examined for each label in each tile category, and the results are plotted in Figure \ref{fig:8}. As shown in Figure \ref{fig:8}, for all three tile activity categories, the $D_a$ declined more in counties with larger population, more confirmed cases, and earlier first-case date. This result indicates a greater recognition and response to the pandemic risks in more populous counties with early confirmed cases.

\section{Discussion}

This study shows the utility of two human activity metrics (the Venables distance ($D_v$) and the activity density ($D_a$)) as leading indicators for the spread speed of COVID-19 in the early stages of the outbreak. These metrics were derived from digital trace data obtained from Mapbox high-resolution temporal-spatial datasets. The results provide statistical evidence regarding the time-lag correlation between these two metrics and the basic reproduction number ($R_0$) in the context of COVID-19. The results regarding the significant leader-follower relationship between human activities and the rate of spread of viral infections could provide valuable implications for authorities to monitor and control the transmission of COVID-19 and future pandemics.  For example, time lag indicates that the spread of virus responds to human activity intensity reduction more quickly than to human agglomeration reduction. Hence, the proposed indicators can be calculated using digital trace telemetry data in near real time to proactively assess the risk of virus spread.
This study has some limitations which need to be improved in future studies. First, the tile activity categorization—social, work, and traffic—is not precise. One tile could be labeled as both social and work. In this study and due to data availability limitations, however, we classified tiles into only one of the three categories. Second, the CDC confirmed-cases data had limitations due to testing availability. In this study, we did not adjust the confirmed case data based on the extent of testing in different counties. A lack of testing in some areas resulted in the underestimation of the total cases.

\section*{Acknowledgements}

The authors would also like to acknowledge that Mapbox provided digital trace telemetry data of human activity. The authors would like to thank Kieran Gupta, Sofia Heisler, Ruggero Tacchi from Mapbox for providing technical support.

\section*{Funding}

This work was supported by several grants including from the United States National Science Foundation RAPID project \#2026814: Urban Resilience to Health Emergencies: “Revealing Latent Epidemic Spread Risks from Population Activity Fluctuations and Collective Sense-making,” and Microsoft AI for Health COVID-19 Grant for cloud computing resources.

\section*{Author contributions statement}

Research design and conceptualization: X. G., C. F., A. M.; Data collection, processing, analysis, and visualization: X. G., C. F., Y. Y., S. L., Q. L; Writing: X. G., A. M.; Reviewing and revising: all authors.

\bibliographystyle{unsrt}  
\bibliography{references}  %%% Remove comment to use the external .bib file (using bibtex).

%\bibliographystyle{plainnat}
%\bibliography{test}


%\begin{thebibliography}{1}

%\bibitem{kour2014real}
%George Kour and Raid Saabne.
%\newblock Real-time segmentation of on-line handwritten arabic script.
%\newblock In {\em Frontiers in Handwriting Recognition (ICFHR), 2014 14th
%  International Conference on}, pages 417--422. IEEE, 2014.

%\bibitem{kour2014fast}
%George Kour and Raid Saabne.
%\newblock Fast classification of handwritten on-line arabic characters.
%\newblock In {\em Soft Computing and Pattern Recognition (SoCPaR), 2014 6th
%  International Conference of}, pages 312--318. IEEE, 2014.

%\bibitem{hadash2018estimate}
%Guy Hadash, Einat Kermany, Boaz Carmeli, Ofer Lavi, George Kour, and Alon
%  Jacovi.
%\newblock Estimate and replace: A novel approach to integrating deep neural
%  networks with existing applications.
%\newblock {\em arXiv preprint arXiv:1804.09028}, 2018.

%\end{thebibliography}

\end{document}